\documentclass[aps,pra,preprint,longbibliography]{revtex4-2}

\usepackage{geometry}
\usepackage{setspace}
\usepackage{graphicx} 
\usepackage{amsmath,amssymb,amsfonts}
\usepackage{appendix}
\usepackage{xcolor}
\usepackage{verbatim}
\makeatletter
\def\@bibdataout@aps{%
  \immediate\write\@bibdataout{%
    @CONTROL{apsrev42Control,
      title = "yes",
      pages = "yes"
    }%
  }%
}
\makeatother

\begin{document}
\title{Non-unitary extension of Grover's search algorithm}
\author{V.N.A. Lula-Rocha}
\affiliation{Departamento de F\'isica, Insituto de Ci\^encias Exatas,\\ 
Universidade Federal de Lavras\\
37200-900 Lavras-MG, Brazil
}
\email{viniciusnonato@gmail.com}

\author{M.A.S. Trindade}
\affiliation{Colegiado de Física, Departamento de Ciências Exatas e da Terra,\\ Universidade do Estado da Bahia\\
Rua Silveira Martins, Cabula,\\ 41150000, Salvador, Bahia, Brazil }
\email{matrindade@uneb.br}

\begin{abstract}
We have developed a non-unitary extension of Grover's search algorithm by changing the hidden geometry of Hilbert space carried by diffusion operator. Our algorithm finds the solution for search problem by performing a unique bigger rotation rather than small rotations in order polynomial times in the size $N$ of search space. We analyze the complexity of implementing the non-unitary operation and we observed that the price paid by performing this rotation is due the normalization. In Kraus operator approach we need $O(N)$ repetition of the algorithm to have a chance of measuring a solution in a post-selection, this is no better than the classical solution. However, the quantum singular value transform in addition with block encoding and Chebyshev polynomial approximation, we got complexity $O(\sqrt{N})$ and reach the Grover's bound with an extra resource of one single qubit, compared with the standard Grover's algorithm.

\bigskip

\end{abstract}
\maketitle
\section{Introduction}

Since the first proposal of quantum computing \cite{Feynman1982,Deutsch1985}, the quantum search algorithm is one of the most fundamental quantum algorithm of the area \cite{MichaelA.Nielsen2010,Portugal2018}.  Developed by Lov Grover \cite{Grover1996} in $1996$, the algorithm now known as Grover's algorithm, finds a solution for the following search problem: we are given an oracle that can evaluate a function $f: \mathbb{Z}_{N}\mapsto \{0,1\}$, such that $f(x_{0})=1$. Without knowing the expression for $f$, we are supposed to find $x_{0}$. In classical computing, this problem is solved by straightforward evaluation of $f$ that requires, in the worst scenario, $N$ evaluations or queries to the oracle. This can be stated \cite{MichaelA.Nielsen2010} that the search problem can be solved in classical computers with complexity order of $O(N)$. In contrast, Grover has showed that in a quantum computer the search problem can be solved by approximately with $\sqrt{N}$ queries to the oracle, or with complexity $O(\sqrt{N})$. This gives to the quantum algorithm a polynomial advantage over the classical one.
 
The Grover's algorithm (GA) together with the quantum algorithms of Deutsch, Deutsch-Josza \cite{Deutsch1992}, Shor \cite{Shor1997}, Simon \cite{Simon1997}, Bernstein-Vazirani \cite{Bernstein1997}, quantum Fourier transform \cite{Cleve1998,Griffiths1996} and quantum phase estimative \cite{Kitaev1995},  is one of the earlier quantum algorithms that demonstrated advantages towards their classical analogues and has become essential to the development of quantum computing itself \cite{MichaelA.Nielsen2010,Portugal2025}, particularly as a motivation to the researchers to reach the actual stage of the area that comprise of noisy intermediate-scale quantum (NISQ) \cite{Abughanem2024,Xie2025,Hai2025,Wei2022} computers with hundreds, or even thousand \cite{Wei2022} of qubits that implement quantum gates, as its name suggest, with a certain controlled noise.

A natural generalization of GA is to allow $f$ to have, not one solution to the problem, but instead, a set $S\subset \mathbb{Z}_{N}$ of solutions. This is what the quantum algorithm of amplitude amplification (AA) \cite{Brassard1997} does. In this case, a single solution is find with approximately $O\left(\sqrt{\frac{N}{|S|}}\right)$ queries to the oracle against $N - |S| +1$ queries in classical algorithms, also establishing quantum advantage. Other generalizations \cite{Tonchev2025,Roy2022,Seidel2023,Biron1999,Byrnes2018,Accardi2000,Gilliam2020,Tulsi2012,Biham2001,Chen2002,Portugal2018,Chowdhury2022,Cardullo2026,Gingrich1999} of GA and AA has been made along the years, for example by allowing arbitrary initial states \cite{Gingrich1999,Biron1999,Cardullo2026} or designed algorithm to find the unitary operation that amplify the probability to find a specific vector component \cite{Accardi2000}. In this context, some recent development includes oracle for distinguished solutions \cite{Tonchev2025}, deterministic search \cite{Roy2022}, oracle design for arbitrary data structures \cite{Seidel2023}, and general phases on oracle and diffusion operator \cite{Chowdhury2022}.

These algorithms can solve problems of search in a unstructured database. This means that elements to be searched are encoded as binary-base integers numbers in $\mathbb{Z}_{N}$, so that can be represented as base states of the quantum processor. The output of the algorithm is then a quantum measurement of the processor that gives the binary representation of the encoded data with high probability. Some of quantum search problems have inspired quantum machine learning models, such as quantum support vector machines \cite{Biamonte2017,Schuld2021} for quantum classification, uses AA to search hyperplanes that separate the classes. We can also find search algorithms applied to quantum cryptography \cite{Jaques2020,Fernandez2024}, post-cryptography \cite{Matthew2017} and optimization problems \cite{Kiktenko2025}. It is worth noting that have been many efforts to analyze different ways and complexities of the Grover's oracle \cite{Yan2022,Stoudenmire2024,Viamontes2005,Liu2024,Amy2017,Jaques2020,Naranjo2024,Abrams1998}. This gave a new brunch of applicability for quantum search algorithms.

A well established result of GA and AA are the proofs that they are the optimal quantum algorithm to solve quantum search \cite{Zalka1999,Boyer1998,Beals2001,Dohotaru2009}. These proofs use as a requirement unitary operations. This is usual, since quantum computing uses quantum systems as carries of information and, due the quantum mechanics, we know that these systems evolves with unitary operators. Nonetheless, recent developments \cite{Daskin2024,Liu2021,Childs2016,Zujev2021,Terashima2005,Heredge2025,An2026,Koukoutsis2025,Daskin2017,Schlimgen2022,Zylberman2025,Gilyen2019,Brearley2025,Chakraborty2024,Gingrich2004,Lin2021,Zhu2025,Castelazo2022,Leadbeater2024} have shown that non-unitary quantum operations can be implemented efficiently in quantum computers. There are several approaches to that aim, but must of them are based in extending the Hilbert space of the problem and a apply a unitary operator in the extended space that acts as a non-unitary operation in the original space. This technique is actually well known in quantum information \cite{Kraus1971,Gaitan2008,Scherer2019,MichaelA.Nielsen2010}, particularly in dealing with quantum noise, where the quantum systems is coupled to environment and a unitary operation that represents the noise model is applied to both systems. The effect of the noise in the quantum system is recovered by tracing out the environment. Non-unitary operations in quantum computing application for real problems have been exploit in producing Gibbs states \cite{Lula-Rocha2023,Nys2024,Lee2022,Selisko2023,Petronilo2023,Zapusek2025,Devulapalli2025}, quantum machine learning \cite{Heredge2025,Lula-Rocha2023,Korutcheva2023,Moro2023,Tuysuz2024,Wiebe2019,Xiao2020,Crawford2018,Huijgen2024,Dixit2022,Minervini2025,Coopmans2024,Salmenpera2023,Coopmans2024a,Bangar2025}, linear differential equations solvers \cite{An2026,Chakraborty2024}, quantum simulation \cite{Daskin2017}, state preparation \cite{Schlimgen2022,Chakraborty2024}, non-unitary-real-space simulation of the diffusion equation \cite{Zylberman2025}, quantum walks \cite{Chakraborty2024,Kadian2021} and quantum algorithms for group convolution \cite{Castelazo2022}.

This fact motivated some GA and AA modifications based on non-unitary operations \cite{Daskin2024,Zujev2021,Daskin2017} looking for new perspective of search problems. Here we will show the diffusion operator used in Grover's and amplitude amplification algorithms is a similarity transformation of a self inverted pseudo-metric tensor $\Lambda$ in the space of quantum operators, therefore a unitary operation. We can generalize these algorithms by allowing $\Lambda$ to be a general $2$-rank tensor and looking for another tensor that enhance the algorithm in the sense of calling less queries to the oracle. We show that is possible to find a new metric tensor such that the algorithm solves the problem of quantum search by only a single call to a non-unitary operation. This operation is a general Grover's operation in the sense that for specific parameters the unitarity is recovered turning into the Grover's operator. 

We analyze computational complexity of the two approaches, the first of them is physical: we study an approximation of the non-unitary operation via Kraus operation. Although this method gives high probability of the non-unitary operation be implemented, the probability of finding a solution to search problem decreases due the normalization process required to guarantee the validity of the Kraus operator. This is compensated by repeating the algorithm in the order of the search space divided by the number of solutions. This, however, is not better than the classical approach. In the second analyzes, we block encoded our non-unitary operation in a larger unitary one \cite{Gilyen2019} and we coherently recover the normalization loss using Chebyshev polynomial approximation. With this method we reach the Grover's bound for a constant error using just a single extra qubit.

Our algorithm is then a new geometric perspective of quantum search problem: we have changed the geometry of Hilbert space which allows us to perform a bigger rotation that rotates the initial state to the space of solution at once. This is in some sense analogue to Abrams' and Lloyd's proposal algorithm for an extension of Grover's search to a nonlinear regime \cite{Abrams1998} that, for sufficient nonlinearity, the oracle is called just once. Algorithms like that could solve $NP$-complete and $\# P$ problems in a polynomial time. There, the strong evidence that quantum mechanics is linear makes nonlinear time evolutions necessary to build such algorithms prohibitive, here non-unitarity ensures that Grover's bound is not violated. However we do not loose the interpretation of a single rotation as the computational complexity is due to implement the non-unitary and not to perform several small rotations. It is worth noting that our algorithm is a non-unitary extension of Grover's algorithm in the sense that solve quantum search problem via a parametrized non-unitary operator which for specific parameters the unitarity and the original algorithm is recovered. This differentiate from another approaches such quantum simulation \cite{MichaelA.Nielsen2010,Zhukov2025,Portugal2018}, approximation of imaginary-time evolution \cite{Suzuki2025} and quantum walks \cite{Portugal2018,Nahimovs2015} that depends on a specific choice of operations such as Hamiltonian or quantum coin to be viewed as a quantum search algorithm. 

This paper is organized in the following way: in Section \ref{theProblem}, we argue that Grover's algorithm can be viewed geometrically by looking diffusion operator as a metric tensor on quantum operation space. In Section \ref{usualGrover}, we fix the notation by reviewing the original Grover's algorithm. Section \ref{metricTensor} is the core of our paper as it is devoted to develop mathematical tools to build our algorithm. In Section \ref{metricGrover}, the algorithm is presented and in Section \ref{implementationCost} we analyze the computational complexity of implementing it. Finally, in Section \ref{conclusions} we share our conclusions by pointing out the strength-ness and limitations of our proposed algorithm.

\section{The geometry behind Grover's algorithm}
\label{theProblem}

The key ingredient of the Grover's algorithm is the diffusion operator whose action upon a state is a reflection of the state \cite{MichaelA.Nielsen2010}
\begin{eqnarray}
|\psi\rangle
&=&
H^{\otimes n}|0\rangle
    \label{theProblem1}
\end{eqnarray}
in the space spanned by $\{|\alpha\rangle, |\beta\rangle\}$, where $|\alpha\rangle$ is the normalized superposition of non solutions of the search problem and $|\beta\rangle$ is the normalized superposition of the solutions. We also have that $n$ is the number of qubits and $H$ is the Hadamard operation
\begin{eqnarray}
H
&=&
\frac{1}{\sqrt{2}}
\left(
\begin{array}{cc}
    1 & 1\\
    1 & -1
\end{array}
\right),
    \label{theProblem2}
\end{eqnarray}
and we have used the abuse of notation $ |0\rangle \equiv |0\rangle^{\otimes n}$.

A convenient way to express the diffusion operator $\mathcal{D}$ \cite{MichaelA.Nielsen2010}  for our proposes here is
\begin{eqnarray}
\mathcal{D}
&=&
H^{\otimes n}
\left[ 
    \:
    2|0\rangle\langle 0 | - I
    \:
\right]
H^{\otimes n}.
    \label{theProblem3}
\end{eqnarray}
Interestingly, when we express the factor $\Lambda = 2|0\rangle\langle 0| -I$ in matrix form, we get a matrix identical to the pseudo-metric tensor for a $2^{n}$ dimensional vector space \cite{Portugal2018}, given by
\begin{eqnarray}
\Lambda
&=&
\left(
\begin{array}{ccccc}
1 & 0 & 0 & \cdots & 0\\
0 & -1 & 0 & \cdots & 0\\
0 & 0 & -1 & \cdots & 0\\
\vdots & \vdots & \vdots & \ddots & \vdots \\
0 & 0 & 0 & \cdots & -1
\end{array}
\right),
    \label{theProblem4}
\end{eqnarray}
where $|0\rangle = ( 1 0 \ldots 0 )^{T}$ and $I$ is the $2^{n}$ dimensional identity matrix. We can then write the diffusion operator as a similarity transformation given by
\begin{eqnarray}
\mathcal{D}
&=&
U
\Lambda
U^{\dagger}.
    \label{theProblem5}
\end{eqnarray}
where $U=H^{\otimes n}$.

In some sense, the first register of first iteration of Grover algorithm
\begin{eqnarray}
|\psi_{1}\rangle|-\rangle
&=&
H^{\otimes n} \Lambda H^{\otimes n} U_{f}
|\psi\rangle|-\rangle
    \label{theProblem5.1}
\end{eqnarray}
lies in the dual space with respect with the transformed metric  tensor $\mathcal{D}$, because
\begin{eqnarray}
|\psi_{1}\rangle
&=&
\mathcal{D} U_{f}
|\psi\rangle|-\rangle,
    \label{theProblem5.2}
\end{eqnarray}
where $|\psi\rangle$ is given by (\ref{theProblem1}), $|-\rangle = H|1\rangle$ and $U_{f}$ is the Grover's oracle, that will see better in next section. Here is important to note that $U_{f}$ is applied over all $n+1$ qubits at once and its action comprises of marking the solutions with a negative phase $e^{i\pi}=-1$.

To fix the notations is important to clarify that action of $\mathcal{D}$ on two vectors $|\phi\rangle$ and $|\phi'\rangle$ is given by $\mathcal{D}(|\phi\rangle,|\phi'\rangle) =\langle \phi'| \mathcal{D} |\phi\rangle$, so that the dual vector $|\phi^{*}\rangle$ of $|\phi\rangle$ is given
\begin{eqnarray}
|\phi^{*}\rangle
&=&
\mathcal{D}(|\phi\rangle, \cdot)
=
\mathcal{D}|\phi\rangle.
    \label{theProblem5.3}
\end{eqnarray}
Note that we are using abuse of notation here when we are giving the meaning of $\mathcal{D}$ both for a linear operator and bilinear operator $\mathcal{D}(\cdot,\cdot)$.

Its also important to keep in mind that the referred dual space here in the context of this paper is in relation to Hilbert space $\mathcal{H}$ endowed with the transformed pseudo-metric tensor $\mathcal{D}$, $(\mathcal{H},\mathcal{D})$. Not to confuse with the bra space, that is dual to the Hilbert space endowed with a Hermitian bilinear form $h$, $(\mathcal{H},h)$. In relation with the second space, the bra $\langle \psi|$ is dual to the ket $|\psi\rangle$ via the action of $h$ in only one vector $|\phi\rangle$, this is $\langle \phi| = h(|\phi\rangle, \cdot)$. We lead the reader to \cite{Greub1975} for deeper issues in linear algebra. In this context, the vector in dual space $|\psi_{1}\rangle$ is dual to the $|\psi\rangle$ with the solutions marked with this negative phase, not dual to $|\psi\rangle$ itself.

We finish this section by rising up the following questions that we claim answer by the end of this presentation: Can we associate $\Lambda$ with a pseudo-metric tensor and exploit search problems for general metrics? This indefinite metric tensor can help us to enhance search problems?

\section{Quantum amplitude amplification algorithm}
\label{usualGrover}

In this section we will fix the notation that we are going to use throughout the paper.

The quantum amplitude amplification algorithm is a generalization of Grover's algorithm. The main question that Grover's algorithm answer is the following: 

Suppose we are given a function $f: \mathbb{Z}_{N} \mapsto \{0,1\}$, where $N = 2^{n}$, the diffusion operator $\mathcal{D}$ and a operation $U_{f}$ that implements
\begin{eqnarray}
U_{f}
|x\rangle|y\rangle
&=&
|x\rangle|y\oplus f(x)\rangle,
    \label{usualGrover1}
\end{eqnarray}
where $\oplus$ is the sum modulo $2$ operation, and a functional quantum computer that can implements the previous operations. The function $f$ is understood to have a single solution to the equation $f(x_{0})=1$. The question is: what is the value of $x_{0}$.

The solution to this question can be obtained with success probability $P_{succ} = 1 - \frac{1}{N}$ of measuring the first register of the state
\begin{eqnarray}
|\psi_{k}\rangle|-\rangle
&=&
\left[
    \left(
        \mathcal{D}\otimes I
    \right)
     U_{f}
\right]^{k}
    H^{\otimes n}\otimes I_{2}
    |0\rangle|-\rangle,
    \label{usualGrover2}
\end{eqnarray}
where 
\begin{eqnarray}
|-\rangle
&=&
\frac{|0\rangle - |1\rangle}{\sqrt{2}},
    \label{usualGrover3}
\end{eqnarray}
and $k$ is of the order of $O(\sqrt{N})$. Its important to note the application of $U_{f}$ on $|x\rangle|-\rangle$, given by
\begin{eqnarray}
U_{f}|x\rangle|-\rangle
&=&
(-1)^{f(x)}|x\rangle|-\rangle.
    \label{usualGrover4}
\end{eqnarray}
In other words, $U_{f}$ inverts the phase of the state $|x_{0}\rangle|-\rangle$, where $x_{0}$ is the solution. This lead to the definition of a bidimensional space spanned by the vectors $\{|\alpha\rangle|-\rangle,|\beta\rangle\|-\rangle\}$, where $|\alpha\rangle =\frac{1}{\sqrt{N}}\sum_{x\neq x_{0}}|x\rangle$ is the superposition of the non-solution states, this is the $|x\rangle$ from which $f(x)=0$ and $|\beta\rangle = \frac{1}{\sqrt{N}}|x_{0}\rangle$. Since $|-\rangle$ is unchanged throughout the whole algorithm, we will omit the second register from the expressions as well as the identities in (\ref{usualGrover2}) that act on it. 

The quantum amplitude amplification is the generalization of Grover's algorithm when $f$ is allowed to have $M$ solutions. In this case, the vectors of the bi-dimensional space has to be redefined as
\begin{eqnarray}
|\alpha\rangle
&=&
\frac{1}{\sqrt{N-M}}
\sum_{m\not\in S}
|m\rangle,
\label{usualGrover5}
\\
|\beta\rangle
&=&
\frac{1}{\sqrt{M}}
\sum_{m\in S}
|m\rangle.
    \label{usualGrover6}
\end{eqnarray}
where $S$ is the set of solutions of $f$. In quantum amplitude amplification \cite{MichaelA.Nielsen2010}, $k$ is of the order of $O\left(\sqrt{\frac{N}{M}}\right)$ and the probability of success is changed to $P_{succ}=1 - \frac{M}{N}$.

\subsection{Iterations of quantum search algorithms}

For our proposes here, is interesting to show details the action of the iterations of (\ref{usualGrover2}). Lets observe that the state $|\psi\rangle$ can be written in terms of $|\alpha\rangle$ and $|\beta\rangle$:
\begin{eqnarray}
|\psi\rangle
&=&
 H^{\otimes}|0\rangle
=
\frac{1}{\sqrt{N}}
\sum_{x=0}^{N-1}
|x\rangle.
    \label{iterationUsualGrover1}
\end{eqnarray}
The sum in (\ref{iterationUsualGrover1}) can be split into two parts, the first sum where $x$ are not in $S$ and the second that are in $S$, as
\begin{eqnarray}
|\psi\rangle
&=&
\cos(\theta/2)
|\alpha\rangle
+
\sin(\theta/2)
|\beta\rangle,
    \label{iterationUsualGrover4}
\end{eqnarray}
where $\cos(\theta/2)=\sqrt{\frac{N-M}{M}}$ and $|\alpha\rangle$ and $|\beta\rangle$ are given by (\ref{usualGrover5}) and (\ref{usualGrover6}).

The diffusion operator can be written in terms of $|\psi\rangle$ as
\begin{eqnarray}
\mathcal{D}
&=&
2|\psi\rangle\langle \psi| - I.
    \label{iterationUsualGrover5}
\end{eqnarray}
We substitute (\ref{iterationUsualGrover4}) in (\ref{iterationUsualGrover5}) and use the completeness relation of the set $\{|\alpha\rangle,|\beta\rangle\}$, this is, $|\alpha\rangle\langle \alpha| + |\beta\rangle\langle\beta| = I$, then
\begin{eqnarray}
\mathcal{D}
&=&
2\cos^{2}(\theta/2)|\alpha\rangle\langle \alpha|
+
2\sin^{2}(\theta)|\beta\rangle\langle \beta|
+
2\sin(\theta/2)\cos(\theta/2)
\left(
    |\alpha\rangle\langle \beta|
    +
    |\beta\rangle\langle \alpha|
\right)
\nonumber
\\
&&
- |\alpha\rangle\langle \alpha| - |\beta\rangle\langle\beta|
    \label{iterationUsualGrover6}
\end{eqnarray}
One can easily show, using the fundamental trigonometric relations, that
\begin{eqnarray}
\mathcal{D}
&=&
\cos(\theta)|\alpha\rangle\langle\alpha|
-
\cos(\theta)|\beta\rangle\langle\beta|
+
\sin(\theta)
\left(
    |\alpha\rangle\langle \beta|
    +
    |\beta\rangle\langle \alpha|
\right).
    \label{iterationUsualGrover8}
\end{eqnarray}
The expression for the first iteration of (\ref{usualGrover2}) is found by applying $\mathcal{D}U_{f}$, with $\mathcal{D}$ written in the form of (\ref{iterationUsualGrover8}), on the initial state $|\psi_{0}\rangle = H^{\otimes n}|0\rangle$. This is
\begin{eqnarray}
|\psi_{1}\rangle
&=&
\left[
\cos(\theta)|\alpha\rangle\langle\alpha|
-
\cos(\theta)|\beta\rangle\langle\beta|
+
\sin(\theta)
\left(
    |\alpha\rangle\langle \beta|
    +
    |\beta\rangle\langle \alpha|
\right)
\right]
\left[
    \cos(\theta/2)|\alpha\rangle 
    -
    \sin(\theta/2)|\beta\rangle
\right].
\nonumber
\\
    \label{iterationUsualGrover9}
\end{eqnarray}
Here we note the application of $U_{f}$ in $|\psi_{0}\rangle$, inverts the phase of the solution states $|\beta\rangle$. This minus sign is fundamental for us to express the first iteration as rotation by $\theta$ in the initial state in $\{|\alpha\rangle, |\beta\rangle\}$ plane. 

By distributing (\ref{iterationUsualGrover9}) we find
\begin{eqnarray}
|\psi_{1}\rangle
&=&
\left(
\cos(\theta)\cos(\theta/2)
-
\sin(\theta)\sin(\theta/2)
\right)
|\alpha\rangle
+
\left(
\cos(\theta)\sin(\theta/2)
+
\sin(\theta)\cos(\theta/2)
\right)
|\beta\rangle
\nonumber\\
    \label{iterationUsualGrover10}
\end{eqnarray}
The above expression show us that a single iteration of $\mathcal{D}U_{f}$ on the initial state adds an angle $\theta$ to it
because
$
\cos(\frac{\theta}{2} + \theta)
=
\cos(\theta)\cos(\theta/2)
-
\sin(\theta)\sin(\theta/2)
$
and
$
\sin(\frac{\theta}{2} + \theta)
=
\cos(\theta)\sin(\theta/2)
+
\sin(\theta)\cos(\theta/2)
$. Using these trigonometric relations write
\begin{eqnarray}
|\psi_{1}\rangle
&=&
\cos(\theta/2 + \theta)|\alpha\rangle
+
\sin(\theta/2 + \theta)|\beta\rangle.
    \label{iterationUsualGrover11}
\end{eqnarray}
For $k$ iteration this mechanism repeats and we have
\begin{eqnarray}
|\psi_{k}\rangle
&=&
\cos(\theta/2 + k\theta)|\alpha\rangle
+
\sin(\theta/2 + k\theta)|\beta\rangle.
    \label{iterationUsualGrover12}
\end{eqnarray}
From (\ref{iterationUsualGrover12}) we can obtain the order of $k$ to approximate the probability $\sin^{2}(\theta/2 + k\theta)$ to $1$ and we find $k=O\left(\sqrt{\frac{N}{M}}\right)$ and the probability of success of $P_{succ}=1-\frac{N}{M}$. For this is necessary to observe that $\theta/2$ can be considered as a small angle such that $\cos(\theta/2)\approx 1$ and $\sin(\theta/2)\approx \theta/2$. Our interest of showing this calculation is to note that the angle increasing is possible due the last term of (\ref{iterationUsualGrover8}). We will show that the view of diffusion operator as a transformed indefinite metric tensor allows us to manipulate those expressions and change the rotation angle by our convenience. This will be made in the following sections.

\section{Search problems from geometric perspective}
\label{metricTensor}

In this section, we will consider a general $2$-tensor $g$ thought to be induced by a general correlation $\tau: \mathcal{H}\mapsto \mathcal{H}^{*}$, where $\mathcal{H}^{*}$ is the dual space of $\mathcal{H}$. In this fashion, $g$ is defined as $g(|\phi\rangle,|\phi'\rangle)=\tau(|\phi\rangle)(|\phi'\rangle)$. We will not make any symmetry, linearity or sesquilinearity so that we are free to chose these features latter depending on our convenience. In braket notation we have
\begin{eqnarray}
g
&=&
\sum_{\ell, k = 0}^{N-1}
g_{\ell k}
|\ell\rangle\langle k|,
    \label{metricTensor1}
\end{eqnarray}
where $g_{\ell, k}$ are complex numbers.

\subsection{On a new vectors}

In order to perform the calculation of a generalization of the diffusion operator $\mathcal{D}_{g}$ for a indefinite metric $g$ as we did in the previous sections, we need to generalize the states $|\alpha\rangle$ and $|\beta\rangle$. In this sense, lets analyze the action of the Hadamard transform $H^{\otimes n}$ on a state of computational basis $|\ell\rangle$,
\begin{eqnarray}
H^{\otimes n}
|\ell\rangle
&=&
\frac{1}{\sqrt{N}}
\sum_{m=0}^{N-1}
(-1)^{\ell\cdot m}
|m\rangle,
    \label{metricTensor2}
\end{eqnarray}
where $\ell\cdot m = \ell_{1}m_{1} + \ell_{2}m_{2} + \ldots + \ell_{n}m_{n}$.

Following the same approach to obtain (\ref{iterationUsualGrover4}), we define
\begin{eqnarray}
|\alpha_{\ell}\rangle
&=&
\frac{1}{\sqrt{N-M}}
\sum_{m\not\in S}
(-1)^{\ell\cdot m}
|m\rangle,
    \label{metricTensor3}
\\
|\beta_{\ell}\rangle
&=&
\frac{1}{\sqrt{M}}
\sum_{m\in S}
(-1)^{\ell\cdot m}
|m\rangle,
    \label{metricTensor4}
\end{eqnarray}
We observe that $|\alpha_{\ell}\rangle$ and $|\beta_{\ell}\rangle$ is still a normalized superposition of non-solutions and solutions of $f(m)=1$, respectively. Nonetheless they do not form a complete and orthonormal basis for our bidimensional space. Actually we have the relations
\begin{eqnarray}
\langle\alpha_{\ell}|\beta_{k}\rangle
&=&
\langle\beta_{k}|\alpha_{\ell}\rangle
=
0;
    \label{metricTensor5}
\\
\langle\alpha_{\ell}|\alpha_{\ell}\rangle
&=&
\langle\beta_{\ell}|\beta_{\ell}\rangle
=
1;
    \label{metricTensor6}
\\
\langle\alpha_{\ell}|\alpha_{k}\rangle
&=&
\langle\beta_{\ell}|\beta_{k}\rangle
\neq
\delta_{\ell k}, 
\qquad
\ell \neq k.
    \label{metricTensor6}
\end{eqnarray}
With this notation, we can express (\ref{metricTensor2}) as
\begin{eqnarray}
H^{\otimes n}|\ell\rangle
&=&
\cos(\theta/2)|\alpha_{\ell}\rangle
+
\sin(\theta/2)|\beta_{\ell}\rangle.
    \label{metricTensor7}
\end{eqnarray}
Observe that if $\ell=0$, we get $|\alpha_{0}\rangle=|\alpha\rangle$, $|\beta_{0}\rangle=|\beta\rangle$ and equation (\ref{metricTensor7}) turns to (\ref{iterationUsualGrover4}). As the Hadamard transform $H^{\otimes n}$ is its own inverse, if we apply it in both sides of (\ref{metricTensor7}) we get $|\ell\rangle$ in terms of $|\alpha_{\ell}\rangle$ and $|\beta_{\ell}\rangle$ as
\begin{eqnarray}
|\ell\rangle
&=&
\frac{1}{\sqrt{N}}
\sum_{m=0}^{N-1}
(-1)^{\ell\cdot m}
\left[
    (1 + \tan(\theta/2))|\alpha_{m}\rangle
    +
    (1 + \cot(\theta/2))|\beta_{m}\rangle
\right].
    \label{metricTensor8}
\end{eqnarray}

\subsection{On a generalized diffusion operator}

The new set of vectors $\{|\alpha_{\ell}\rangle, |\beta_{k}\rangle\}$ allows us to calculate the generalized diffusion operator $\mathcal{D}_{g}$ in the lines of the previous sections,
\begin{eqnarray}
\mathcal{D}_{g}
&=&
H^{\otimes n}gH^{\otimes n}
\nonumber
\\
&=&
\sum_{\ell,k=0}^{N-1}
g_{\ell k}
\left[
    \cos(\theta/2)|\alpha_{\ell}\rangle 
    + 
    \sin(\theta/2)|\beta_{\ell}\rangle
\right]
\left[
    \cos(\theta/2)\langle\alpha_{k}| 
    + 
    \sin(\theta/2)\langle\beta_{k}|
\right]
\nonumber
\\
&=&
\sum_{\ell,k=0}^{N-1}
g_{\ell k}
\left[
    \cos^{2}(\theta/2)|\alpha_{\ell}\rangle\langle\alpha_{k}|
    +
    \sin^{2}(\theta/2)|\beta_{\ell}\rangle\langle\beta_{k}|
\right.
\nonumber\\
&&
\left.
+
\sin(\theta/2)\cos(\theta/2)
\left(
    |\alpha_{\ell}\rangle\langle\beta_{k}|
    +
    |\beta_{\ell}\rangle\langle\alpha_{k}|
\right)
\right].
    \label{metricTensor9}
\end{eqnarray}
Since we know from previous calculations that $|\alpha_{0}\rangle=|\alpha\rangle$ and $|\beta_{0}\rangle=|\beta\rangle$, it is interesting separete the sum (\ref{metricTensor9}) from the term corresponding to $\ell,k=0$. We get
\begin{eqnarray}
\mathcal{D}_{g}
&=&
g_{00}
\mathcal{D}
+
\sum_{\ell,k=1}^{N-1}
g_{\ell k}
\left[
    \cos^{2}(\theta/2)|\alpha_{\ell}\rangle\langle\alpha_{k}|
    +
    \sin^{2}(\theta/2)|\beta_{\ell}\rangle\langle\beta_{k}|
\right.
\nonumber\\
&&
\left.
+
\sin(\theta/2)\cos(\theta/2)
\left(
    |\alpha_{\ell}\rangle\langle\beta_{k}|
    +
    |\beta_{\ell}\rangle\langle\alpha_{k}|
\right)
\right].
    \label{metricTensor10}
\end{eqnarray}
We can also explicit the diagonal and non-diagonal contributions,
\begin{eqnarray}
\mathcal{D}_{g}
&=&
g_{00}
\mathcal{D}
+
\sum_{\ell=1}^{N-1}
g_{\ell \ell}
\left[
    \cos^{2}(\theta/2)|\alpha_{\ell}\rangle\langle\alpha_{\ell}|
    +
    \sin^{2}(\theta/2)|\beta_{\ell}\rangle\langle\beta_{\ell}|
\right.
\nonumber\\
&&
\left.
+
\sin(\theta/2)\cos(\theta/2)
\left(
    |\alpha_{\ell}\rangle\langle\beta_{\ell}|
    +
    |\beta_{\ell}\rangle\langle\alpha_{\ell}|
\right)
\right]
\nonumber
\\
\nonumber
\\
&&
+
\sum_{\substack{\ell\neq k \\ \ell,k\neq 0}}
g_{\ell k}
\left[
    \cos^{2}(\theta/2)|\alpha_{\ell}\rangle\langle\alpha_{k}|
    +
    \sin^{2}(\theta/2)|\beta_{\ell}\rangle\langle\beta_{\ell}|
\right.
\nonumber\\
&&
\left.
+
\sin(\theta/2)\cos(\theta/2)
\left(
    |\alpha_{\ell}\rangle\langle\beta_{k}|
    +
    |\beta_{\ell}\rangle\langle\alpha_{k}|
\right)
\right].
    \label{metricTensor11}
\end{eqnarray}

\subsection{On a metric quantum search algorithm}

We can now propose a new algorithm for quantum search based on quantum amplitude amplification. The proposal is to associate $\mathcal{D}\mapsto \mathcal{D}_{g}$ in (\ref{usualGrover2}). In the language of Section \ref{theProblem}, the first iteration of the algorithm leads the initial state to the dual space associated with the general $2$-rank tensor $g$ rather than the pseudo-metric tensor $\Lambda$. We will get a quantum amplitude amplification like algorithm with $N^{2}$ free parameters $g_{\ell,k}$ that we claim help us to enhance the quantum search. For the first iteration of (\ref{usualGrover2}), this association is
\begin{eqnarray}
|\psi_{1}\rangle
&=&
\mathcal{D}U_{f}
|\psi_{0}\rangle
\quad
\longmapsto
|\psi_{g,1}\rangle
=
\mathcal{D}_{g}U_{f}
|\psi_{0}\rangle,
    \label{metricGrover1}
\end{eqnarray}
where $|\psi_{0}\rangle$ is the initial state $|\psi_{0}\rangle = H^{\otimes n}|0\rangle$. Using (\ref{metricTensor10}) to apply $\mathcal{D}_{g}$ in $U_{f}|\psi_{0}\rangle= \cos^{2}(\theta/2)|\alpha\rangle - \sin^{2}(\theta/2)|\beta\rangle$, we have
\begin{eqnarray}
|\psi_{g,1}\rangle
&=&
g_{00}\mathcal{D}U_{f}|\psi_{0}\rangle
+
\sum_{\ell,k=1}^{N-1}
g_{\ell k}
\left[
    \cos^{2}(\theta/2)|\alpha_{\ell}\rangle\langle\alpha_{k}|
    +
    \sin^{2}(\theta/2)|\beta_{\ell}\rangle\langle\beta_{k}|
\right.
\nonumber\\
&&
\left.
+
\sin(\theta/2)\cos(\theta/2)
\left(
    |\alpha_{\ell}\rangle\langle\beta_{k}|
    +
    |\beta_{\ell}\rangle\langle\alpha_{k}|
\right)
\right]
U_{f}|\psi_{0}\rangle
\nonumber
\\
\nonumber
\\
&=&
g_{0 0}
\left[
    \cos^{2}(\theta/2)|\alpha\rangle\langle\alpha|
    +
    \sin^{2}(\theta/2)|\beta\rangle\langle\beta|
\right.
\nonumber
\\
\nonumber
\\
&&
\left.
    +
    \sin(\theta/2)\cos(\theta/2)
    \left(
        |\alpha\rangle\langle\beta| + |\beta\rangle\langle\alpha|
    \right)
\right]
\left[
    \cos(\theta/2)|\alpha\rangle 
    -
    \sin(\theta/2)|\beta\rangle
\right]
\nonumber
\\
\nonumber
\\
&&
+
\sum_{\ell,k=1}^{N-1}
g_{\ell k}
\left[
    \cos^{2}(\theta/2)|\alpha_{\ell}\rangle\langle\alpha_{k}|
    +
    \sin^{2}(\theta/2)|\beta_{\ell}\rangle\langle\beta_{k}|
\right.
\nonumber\\
&&
\left.
+
\sin(\theta/2)\cos(\theta/2)
\left(
    |\alpha_{\ell}\rangle\langle\beta_{k}|
    +
    |\beta_{\ell}\rangle\langle\alpha_{k}|
\right)
\right]
\left[
    \cos(\theta/2)|\alpha\rangle 
    -
    \sin(\theta/2)|\beta\rangle
\right].
\nonumber
\\
    \label{metricGrover2}
\end{eqnarray}
By distributing the multiplication and putting together the coefficients of the same kets, we get
\begin{eqnarray}
|\psi_{g,1}\rangle
&=&
g_{00}
\left[
    \cos(\theta/2)
    \left(
        \cos^{2}(\theta/2) - \sin^{2}(\theta/2)
    \right)
    |\alpha\rangle
    +
    \sin(\theta/2)
    \left(
        \cos^{2}(\theta/2) - \sin^{2}(\theta/2)
    \right)
    |\beta\rangle
\right]
\nonumber
\\
&&
+
\sum_{\ell,k=1}^{N-1}
g_{\ell k}
\left[
    \cos(\theta/2)
    \left(
        \cos^{2}(\theta/2)\langle\alpha_{k}|\alpha\rangle
        -
        \sin^{2}(\theta/2)\langle\beta_{k}|\beta\rangle
    \right)
    |\alpha_{\ell}\rangle
\right.
\nonumber
\\
\nonumber
\\
&&
\left.
+
\sin(\theta/2)
\left(
    \cos^{2}(\theta/2)\langle\alpha_{k}|\alpha\rangle
    -
    \sin^{2}(\theta/2)\langle\beta_{k}|\beta\rangle
\right)
|\beta_{\ell}\rangle
\right]
    \label{metricGrover3}
\end{eqnarray}
Now we can separate the sum into diagonal and non-diagonal terms as we made above, then
\begin{eqnarray}
|\psi_{g,1}\rangle
&=&
g_{00}
\left[
    \cos(\theta/2)
    \left(
        \cos^{2}(\theta/2) - \sin^{2}(\theta/2)
    \right)
    |\alpha\rangle
    +
    \sin(\theta/2)
    \left(
        \cos^{2}(\theta/2) - \sin^{2}(\theta/2)
    \right)
    |\beta\rangle
\right]
\nonumber
\\
&&
+
\sum_{\ell=1}^{N-1}
g_{\ell \ell}
\left[
    \cos(\theta/2)
    \left(
        \cos^{2}(\theta/2)\langle\alpha_{\ell}|\alpha\rangle
        -
        \sin^{2}(\theta/2)\langle\beta_{\ell}|\beta\rangle
    \right)
    |\alpha_{\ell}\rangle
\right.
\nonumber
\\
\nonumber
\\
&&
\left.
+
\sin(\theta/2)
\left(
    \cos^{2}(\theta/2)\langle\alpha_{\ell}|\alpha\rangle
    -
    \sin^{2}(\theta/2)\langle\beta_{\ell}|\beta\rangle
\right)
|\beta_{\ell}\rangle
\right]
\nonumber
\\
\nonumber
\\
&&
+
\sum_{\substack{\ell\neq k}}
g_{\ell k}
\left[
    \cos(\theta/2)
    \left(
        \cos^{2}(\theta/2)\langle\alpha_{k}|\alpha\rangle
        -
        \sin^{2}(\theta/2)\langle\beta_{k}|\beta\rangle
    \right)
    |\alpha_{\ell}\rangle
\right.
\nonumber
\\
\nonumber
\\
&&
\left.
+
\sin(\theta/2)
\left(
    \cos^{2}(\theta/2)\langle\alpha_{k}|\alpha\rangle
    -
    \sin^{2}(\theta/2)\langle\beta_{k}|\beta\rangle
\right)
|\beta_{\ell}\rangle
\right].
    \label{metricGrover4}
\end{eqnarray}
Observe that equation (\ref{metricGrover4}) is written in terms of $\{|\alpha_{\ell}\rangle, |\beta_{\ell}\rangle\}_{\ell\in Z_{N}}$. It is interesting project this result into the known basis $\{|\alpha\rangle,|\beta\rangle\}$ so we can substitute $g$ for $\Lambda$ of the usual quantum amplitude amplification algorithm in order to compare both formulations and find unknown quantities related to $\langle\alpha_{k}|\alpha\rangle$ and $\langle\beta_{k}|\beta\rangle$. We will use the completeness relation of the later basis
\begin{eqnarray}
|\alpha\rangle\langle\alpha| + |\beta\rangle\langle\beta|
&=&
1,
    \label{metricGrover5}
\end{eqnarray}
to project $|\alpha_{\ell}\rangle$ and $|\beta_{\ell}\rangle$ into
\begin{eqnarray}
|\alpha_{\ell}\rangle
&=&
\langle \alpha|\alpha_{\ell}\rangle|\alpha\rangle,
    \label{metricGrover6}
\\
|\beta_{\ell}\rangle
&=&
\langle \beta|\beta_{\ell}\rangle|\beta\rangle.
    \label{metricGrover7}
\end{eqnarray}
Using (\ref{metricGrover6}) and (\ref{metricGrover7}) in (\ref{metricGrover4}) we obtain
\begin{eqnarray}
|\psi_{g,1}\rangle
&=&
g_{00}
\left[
    \cos(\theta/2)
    \left(
        \cos^{2}(\theta/2) - \sin^{2}(\theta/2)
    \right)
    |\alpha\rangle
    +
    \sin(\theta/2)
    \left(
        \cos^{2}(\theta/2) - \sin^{2}(\theta/2)
    \right)
    |\beta\rangle
\right]
\nonumber
\\
&&
+
\sum_{\ell=1}^{N-1}
g_{\ell \ell}
\left[
    \cos(\theta/2)
    \left(
        \cos^{2}(\theta/2)
        |\langle\alpha_{\ell}|\alpha\rangle|^{2}
        -
        \sin^{2}(\theta/2)
        \langle\beta_{\ell}|\beta\rangle
        \langle\alpha|\alpha_{\ell}\rangle
    \right)
    |\alpha\rangle
\right.
\nonumber
\\
\nonumber
\\
&&
\left.
+
\sin(\theta/2)
\left(
    \cos^{2}(\theta/2)
    \langle\alpha_{\ell}|\alpha\rangle\langle \beta|\beta_{\ell}\rangle
    -
    \sin^{2}(\theta/2)
    |\langle\beta_{\ell}|\beta\rangle|^{2}
\right)
|\beta\rangle
\right]
\nonumber
\\
\nonumber
\\
&&
+
\sum_{\substack{\ell\neq k}}
g_{\ell k}
\left[
    \cos(\theta/2)
    \left(
        \cos^{2}(\theta/2)
        \langle\alpha_{k}|\alpha\rangle
        \langle \alpha|\alpha_{\ell}\rangle
        -
        \sin^{2}(\theta/2)
        \langle\beta_{k}|\beta\rangle
        \langle\alpha|\alpha_{\ell}\rangle
    \right)
    |\alpha\rangle
\right.
\nonumber
\\
\nonumber
\\
&&
\left.
+
\sin(\theta/2)
\left(
    \cos^{2}(\theta/2)\langle\alpha_{k}|\alpha\rangle \langle \beta|\beta_{\ell}\rangle
    -
    \sin^{2}(\theta/2)\langle\beta_{k}|\beta\rangle \langle \beta|\beta_{\ell}\rangle 
\right)
|\beta\rangle
\right].
    \label{metricGrover8}
\end{eqnarray}

\subsection{Usual quantum amplitude amplification with geometric point of view}

In this section we will make an important step towards the understanding of the formulation of the previous sections. We will use the $2$-rank tensor $g=\Lambda$, where $\Lambda$ is defined in Section \ref{theProblem} as
\begin{eqnarray}
\Lambda
&=&
\textrm{diag}
(1,-1,-1,\ldots,-1).
    \label{metricUsualGrover1}
\end{eqnarray}
In this case we will have the usual amplitude amplification but in terms or our $2$-rank tensor approach developed in previous sections. This will help us to find important mathematical relations by comparing (\ref{metricGrover8}) with its analogue using $g=\Lambda$.

We have that the generalized diffusion operator for the pseudo-metric tensor $\Lambda$ coincides with the usual diffusion operator of $\mathcal{D}$ the Grover's algorithm of the equation (\ref{theProblem5}) and is given by
\begin{eqnarray}
\mathcal{D}_{\Lambda}
&=&
\cos^{2}(\theta/2)|\alpha\rangle\langle\alpha|
+
\sin^{2}(\theta/2)|\beta\rangle\langle\beta|
+
\sin(\theta/2)\cos(\theta/2)
\left(
    |\alpha\rangle\langle\beta|
    +
    |\beta\rangle\langle\alpha|
\right)
\nonumber
\\
\nonumber
\\
&&
-
\left[
    \cos^{2}(\theta/2)
    \sum_{\ell=1}^{N-1}
    |\alpha_{\ell}\rangle\langle\alpha_{\ell}|
    +
    \sin^{2}(\theta/2)
    \sum_{\ell=1}^{N-1}
    |\beta_{\ell}\rangle\langle\beta_{\ell}|
\right.
\nonumber\\
&&
\left.
+
\sin(\theta/2)\cos(\theta/2)
\sum_{\ell=1}^{N-1}
\left(
    |\alpha_{\ell}\rangle\langle\beta_{\ell}|
    +
    |\beta_{\ell}\rangle\langle\alpha_{\ell}|
\right)
\right].
    \label{metricUsualGrover2}
\end{eqnarray}

By comparing (\ref{metricUsualGrover2}) with (\ref{iterationUsualGrover6}), we will arrive to the expressions

\begin{eqnarray}
&&
\left[
    \cos^{2}(\theta/2)
    \sum_{\ell=1}^{N-1}
    |\alpha_{\ell}\rangle\langle\alpha_{\ell}|
    +
    \sin^{2}(\theta/2)
    \sum_{\ell=1}^{N-1}
    |\beta_{\ell}\rangle\langle\beta_{\ell}|
+
\sin(\theta/2)\cos(\theta/2)
\sum_{\ell=1}^{N-1}
\left(
    |\alpha_{\ell}\rangle\langle\beta_{\ell}|
    +
    |\beta_{\ell}\rangle\langle\alpha_{\ell}|
\right)
\right]
\nonumber
\\
\nonumber
\\
&=&
-\cos^{2}(\theta/2)|\alpha\rangle\langle\alpha|
-
\sin^{2}(\theta/2)|\beta\rangle\langle\beta|
-
\sin(\theta/2)\cos(\theta/2)
\left(
    |\alpha\rangle\langle\beta|
    +
    |\beta\rangle\langle\alpha|
\right)
\nonumber
\\
\nonumber
\\
&&
+
|\alpha\rangle\langle\alpha| + |\beta\rangle\langle \beta|
\nonumber
\\
\nonumber
\\
&=&
\sin^{2}(\theta/2)|\alpha\rangle\langle\alpha| 
+
\cos^{2}(\theta/2)|\beta\rangle\langle\beta|
-
\sin(\theta/2)\cos(\theta/2)
\left(
    |\alpha\rangle\langle\beta|
    +
    |\beta\rangle\langle\alpha|
\right).
    \label{metricUsualGrover3}
\end{eqnarray}
Now we compare the first and the last line of (\ref{metricUsualGrover3}), and we get the  relations
\begin{eqnarray}
&&
\sum_{\ell=1}^{N-1}
|\alpha_{\ell}\rangle\langle\alpha_{\ell}|
=
\tan^{2}(\theta/2)
|\alpha\rangle\langle\alpha|,
    \label{metricUsualGrover4}
\\
&&
\sum_{\ell=1}^{N-1}
|\beta_{\ell}\rangle\langle\beta_{\ell}|
=
\cot^{2}(\theta/2)
|\beta\rangle\langle\beta|,
    \label{metricUsualGrover5}
\\
&&
\sum_{\ell=1}^{N-1}
\left(
    |\alpha_{\ell}\rangle\langle\beta_{\ell}|
    +
    |\beta_{\ell}\rangle\langle\alpha_{\ell}|
\right)
=
-
\left(
    |\alpha\rangle\langle\beta|
    +
    |\beta\rangle\langle\alpha|
\right).
    \label{metricUsualGrover6}
\end{eqnarray}
The above expressions will be useful for us when it is projected by both sides into $\{|\alpha\rangle,|\beta\rangle\}$ basis, then we get
\begin{eqnarray}
&&
\sum_{\ell=1}^{N-1}
|\langle\alpha|\alpha_{\ell}\rangle|^{2}
=
\tan^{2}(\theta/2),
    \label{metricUsualGrover7}
\\
&&
\sum_{\ell=1}^{N-1}
|\langle\beta|\beta_{\ell}\rangle|^{2}
=
\cot^{2}(\theta/2),
    \label{metricUsualGrover8}
\\
&&
\sum_{\ell=1}^{N-1}   \langle\alpha|\alpha_{\ell}\rangle\langle\beta_{\ell}|\beta\rangle
=
-1.
    \label{metricUsualGrover9}
\end{eqnarray}
This quantities will be important to manipulate expressions like (\ref{metricGrover8}), because we have projected it into the $\{|\alpha\rangle,|\beta\rangle\}$ basis.

\section{Non-unitary generalization of quantum amplitude amplification algorithm}
\label{metricGrover}

Now we have sufficient ingredients to propose new algorithms for quantum amplitude amplification. Let us observe the equation (\ref{metricGrover8}) and equations (\ref{metricUsualGrover7}--\ref{metricUsualGrover9}). We observe that it is possible to use the latter one in (\ref{metricGrover8}) if we consider $g$ as diagonal and $g_{\ell \ell}=\lambda$, where $\lambda$ is any constant complex number, for all $\ell>0$. So we get
\begin{eqnarray}
|\psi_{g,1}\rangle
&=&
g_{00}
\left[
    \cos(\theta/2)
    \cos(\theta)
    |\alpha\rangle
    +
    \sin(\theta/2)
    \cos(\theta)
    |\beta\rangle
\right]
\nonumber
\\
&&
+
\lambda
\left[
    \cos(\theta/2)
    \left(
        \cos^{2}(\theta/2)
        \sum_{\ell=1}^{N-1}
        |\langle\alpha_{\ell}|\alpha\rangle|^{2}
        -
        \sin^{2}(\theta/2)
        \sum_{\ell=1}^{N-1}
        \langle\beta_{\ell}|\beta\rangle
        \langle\alpha|\alpha_{\ell}\rangle
    \right)
    |\alpha\rangle
\right.
\nonumber
\\
\nonumber
\\
&&
\left.
+
\sin(\theta/2)
\left(
    \cos^{2}(\theta/2)
    \sum_{\ell=1}^{N-1}
    \langle\alpha_{\ell}|\alpha\rangle\langle \beta|\beta_{\ell}\rangle
    -
    \sin^{2}(\theta/2)
    \sum_{\ell=1}^{N-1}
    |\langle\beta_{\ell}|\beta\rangle|^{2}
\right)
|\beta\rangle
\right],
\nonumber
\\
    \label{metricGrover10}
\end{eqnarray}
where we use 
$\cos^{2}(\theta/2) - \sin^{2}(\theta/2) = \cos(\theta)$. Using (\ref{metricUsualGrover7}--\ref{metricUsualGrover9}) we arrive at
\begin{eqnarray}
|\psi_{g,1}\rangle
&=&
g_{00}
\left[
    \cos(\theta/2)
    \cos(\theta)
    |\alpha\rangle
    +
    \sin(\theta/2)
    \cos(\theta)
    |\beta\rangle
\right]
\nonumber
\\
&&
+
\lambda
\left[
    \cos(\theta/2)
    \left(
        \cos^{2}(\theta/2)
        \tan^{2}(\theta/2)
        +
        \sin^{2}(\theta/2)
    \right)
    |\alpha\rangle
\right.
\nonumber
\\
\nonumber
\\
&&
\left.
-
\sin(\theta/2)
\left(
    \cos^{2}(\theta/2)
    +
    \sin^{2}(\theta/2)
    \cot^{2}(\theta/2)
\right)
|\beta\rangle
\right],
\nonumber
\\
    \label{metricGrover11}
\end{eqnarray}
Using basic trigonometric relations on above, the expression (\ref{metricGrover11}) can be rewritten as
\begin{eqnarray}
|\psi_{g,1}\rangle
&=&
g_{00}
\left[
    \cos(\theta/2)
    \cos(\theta)
    |\alpha\rangle
    +
    \sin(\theta/2)
    \cos(\theta)
    |\beta\rangle
\right]
\nonumber
\\
&&
+
\lambda
\left[
    \sin(\theta)\sin(\theta/2)|\alpha\rangle
    -
    \sin(\theta)\cos(\theta/2)|\beta\rangle
\right].
    \label{metricGrover12}
\end{eqnarray}
This is the same as
\begin{eqnarray}
|\psi_{g,1}\rangle
&=&
\left[
    g_{00} 
    \cos(\theta/2)\cos(\theta) 
    +
    \lambda
    \sin(\theta)\sin(\theta/2)
\right]
|\alpha\rangle
\nonumber
\\
&&
+
\left[
    g_{00}
    \sin(\theta/2)\cos(\theta)
    -
    \lambda
    \sin(\theta)\cos(\theta/2)
\right]
|\beta\rangle.
    \label{metricGrover13}
\end{eqnarray}
Now we understand the meaning of $\Lambda=\textrm{diag}(1,-1,-1,\ldots,-1)$. This mean that for $g=\Lambda$, we have $g_{00}=1$ and $\lambda=-1$, so that
\begin{eqnarray}
|\psi_{\Lambda,1}\rangle
&=&
\left[
    \cos(\theta/2)\cos(\theta) 
    -
    \sin(\theta)\sin(\theta/2)
\right]
|\alpha\rangle
+
\left[
    \sin(\theta/2)\cos(\theta)
    +
    \sin(\theta)\cos(\theta/2)
\right]
|\beta\rangle.
\nonumber
\\
    \label{metricGrover14}
\end{eqnarray}
That is exactly 
\begin{eqnarray}
|\psi_{\Lambda,1}\rangle
&=&
\cos(\theta/2+\theta)|\alpha\rangle
+
\sin(\theta/2+\theta)|\beta\rangle.
    \label{metricGrover15}
\end{eqnarray}
This is the same expression as (\ref{iterationUsualGrover11}). As we see, in this fashion we need the order of $\sqrt{\frac{N}{M}}$ iterations of $\mathcal{D}_{\Lambda}U_{f}$ on the initial state. But here we have in hand two free parameters that we can use to try to enhance the algorithm. Suppose we reduce to one free parameter by choosing
\begin{eqnarray}
g_{0,0}
&=&
\frac{\cos(\phi)}{\cos(\theta)};
\qquad\qquad
\lambda
=
-
\frac{\sin(\phi)}{\sin(\theta)}.
    \label{metricGrover16}
\end{eqnarray}
Substituting this choice in (\ref{metricGrover13}) we get
\begin{eqnarray}
|\psi_{g,1}(\phi)\rangle
&=&
\cos(\theta/2+\phi)|\alpha\rangle
+
\sin(\theta/2+\phi)|\beta\rangle,
    \label{metricGrover17}
\end{eqnarray}
where we explicited the $\phi$-dependence on the state $|\psi_{g,1}\rangle$. For example, for $\phi=k\theta$ we have the state resulting of the $k$th iteration of the usual quantum amplitude amplification algorithm:
\begin{eqnarray}
|\psi_{g,1}(k\theta)\rangle
&=&
\cos(\theta/2+k\theta)|\alpha\rangle
+
\sin(\theta/2+k\theta)|\beta\rangle,
    \label{metricGrover18}
\end{eqnarray}
if we consider $k$ as an integer. But we have made any consideration of $\phi$ nor $k$. We can choose $\phi=\pi/2-\theta/2$, or equivalently $k= \frac{\pi}{2\theta}-\frac{1}{2}$, and get
\begin{eqnarray}
|\psi_{g,1}(\pi/2-\theta/2)\rangle
&=&
\cos(\pi/2)|\alpha\rangle
+
\sin(\pi/2)|\beta\rangle
    \label{metricGrover19}
\end{eqnarray}
or
\begin{eqnarray}
|\psi_{g,1}(\pi/2-\theta/2)\rangle
&=&
|\beta\rangle.
    \label{metricGrover20}
\end{eqnarray}
What we developed here is a method to create a linear transformation $\mathcal{D}_{g}U_{f}$ capable to rotate the initial state $|\psi_{0}\rangle$ into the subspace of solutions with only a single application. It is worth to note that by measuring $|\psi_{g,1}(\pi/2-\theta/2)\rangle$ we have probability $P(x_i)=1$ of  obtaining a solution $x_{i}\in S$ of $f(x)=1$. Nonetheless, it is in practice prohibitive to appy $\mathcal{D}_{g}$ on a quantum computer since $g=\textrm{diag}(\cos(\phi)/\cos(\theta),-\sin(\phi)/\sin(\theta),\ldots,-\sin(\phi)/\sin(\theta))$ is not unitary.

\section{Implementations of non-unitary operations on quantum computer}
\label{implementationCost}

As we see, we can find a solution to Grover's problem with only a call to the oracle $U_{f}$. This result seems to violate the proof of optimality of Grover's algorithm \cite{Beals2001,Boyer1998,Dohotaru2009,Zalka1999} that state that no quantum algorithm can solve search problem with less than $O\left(\sqrt{\frac{N}{M}}\right)$ calls to the oracle, where $N$ is the search space size and $M$ is the number of solutions. Our algorithm solves this problem with complexity $O(1)$, but only mathematically, because quantum computer cannot perform non-unitary operations. Although these proofs requires  unitary operation, if we want to implement our algorithm in a quantum computer it is needed to use techniques to implement non-unitary circuits.

Most of these techniques comprises of dilatation, this is to enlarge the problem's Hilbert space and perform a unitary operation to the enlarged space so that it can reproduce the non-unitary operation in the original space. There are other possibilities, like imaginary time evolution \cite{Lin2021}, Thermofield Dynamics-like Bogoliubov transformation \cite{Lula-Rocha2023}, exponentiation of anti-Hermitian operators \cite{Gingrich2004}, linear combination of unitaries \cite{Chakraborty2024} and others \cite{Daskin2017,An2026,Schlimgen2022,Zylberman2025,Zhu2025,Daskin2024,Koukoutsis2025}.

\subsection{Kraus operator approach}

Here we will not focous in a specific implementation, but in the general cost to implement the non-unitary $D_g U_{f}$. First of all, we will note that any valid Kraus $K$ operator must be positive obey $K^{\dagger}K \leq I$. Our $D_{g}$ is not positive, but it is possible to perform a singular value transformation to write $D_{g}=U D'_{g} V$, where $U$  is the $2^{n}\times 2^{n}$ identity matrix and $V=\textrm{diag}(1, -1, -1, \ldots, -1)$ and
\begin{eqnarray}
D'_{g}
=
\left(
    \begin{matrix}
        \frac{\cos\phi}{\cos\theta} & 0           & \cdots & 0\\
        0          & \frac{\sin\phi}{\sin\theta} & \cdots & 0\\
        \vdots     &  \vdots     & \ddots & 0\\
        0          &   0         & \cdots & \frac{\sin\phi}{\sin\theta}
    \end{matrix}
\right).
    \label{implementationCost0}
\end{eqnarray}

Since $U_{f}$, $U$ and $V$ are unitary, then our problem lays in $D'_{g}$ and we have
\begin{eqnarray}
(D_{g}U_{f})^{\dagger}D_{g}U_{f} 
&=&
D_{g}^{\dagger}D_{g}
=
D_{g}^{\prime\dagger}D'_{g}
=
\left(
    \begin{matrix}
        g_{00}^{2} & 0           & \cdots & 0\\
        0          & \lambda^{2} & \cdots & 0\\
        \vdots     &  \vdots     & \ddots & 0\\
        0          &   0         & \cdots & \lambda^{2}
    \end{matrix}
\right),
    \label{implementationCost1}
\end{eqnarray}
with $g_{00}=\frac{\cos\phi}{\cos\theta}$ and $\lambda=\frac{\sin\phi}{\sin\theta}$.  In the case that $\phi=\theta$ we will have $D_{g}U_{f}$ unitary and equals to one Grover's iteration. Supose we want to outperform Grover's algorithm. Then, since for any advantage of rotation for our algorithm, we need $\theta < \phi < \pi/2$. This will always lead to $\sin\phi > \sin\theta$ and $\lambda > 1$. We can conclude that, if we want to outperform Grover's algorithm, we will have
\begin{eqnarray}
D_{g}^{\prime\dagger}D'_{g}
&>&
I.
    \label{implementationCost2}
\end{eqnarray}

To transform $D_{g}^{\prime}$ in a valid Kraus operator we must renormalize it by defining the following operation \cite{Gilyen2019}
\begin{eqnarray}
\mathcal{D}_{g}
&=&
\sqrt{p}D_{g}^{\prime},
    \label{implementationCost3}
\end{eqnarray}
so that
\begin{eqnarray}
\mathcal{D}_{g}^{\prime\dagger}\mathcal{D}_{g}^{\prime}
&=&
pD_{g}^{\prime\dagger}D_{g}^{\prime}
\leq
I.
    \label{implementationCost4}
\end{eqnarray}
This requires that $p \leq \frac{1}{||D_{g}^{\prime}||^{2}}$, where $||\cdot||$ is the spectral norm. To outperform Grover's algorithm the equality in (\ref{implementationCost4}) will never holds, so $\mathcal{D}_{g}$ would not be complete positive nor trace preserving operation. The consequence of this is that $D_{g}$ cannot be implemented in a quantum computer, but only can be approximated by the quantum operator
\begin{eqnarray}
\mathcal{E}(\rho)
&=&
\mathcal{D}_{g}\rho \mathcal{D}_{g}^{\dagger}
+
F\rho F^{\dagger},
    \label{implementationCost5}
\end{eqnarray}
with $F=\sqrt{I - \mathcal{D}_{g}\mathcal{D}_{g}^{\dagger}}$. Since we are interested in measuring the state $\sigma=\mathcal{D}_{g}\rho \mathcal{D}_{g}^{\dagger}$, we must calculate the probability
\begin{eqnarray}
P(\sigma)
&=&
\textrm{Tr}
(\mathcal{D}_{g}^{\prime}\rho \mathcal{D}_{g}^{\prime\dagger})
=
p
\textrm{Tr}
(D_{g}^{\prime}\rho D_{g}^{\prime\dagger}).
    \label{implementationCost6}
\end{eqnarray}
By (\ref{implementationCost4}) we have that
\begin{eqnarray}
P(\sigma)
&\leq&
\frac{1}{||D_{g}^{\prime}||^{2}}
\textrm{Tr}
\left(
D_{g}^{\prime}
\rho
D_{g}^{\prime\dagger}
\right),
    \label{implementationCost7}
\end{eqnarray}
that represents the probability of success of applying the non-unitary $D_{g}^{\prime}$ by using the Kraus operator $\mathcal{D}_{g}$.

If we specify that $\rho$ is the oracle $U_{f}$ applied to the initial state of the algorithm, this is $\rho = V U_{f}|\psi\rangle\langle\psi|U_{f}^{\dagger}V^{\dagger}$ with $|\psi\rangle$ given by (\ref{theProblem1}), then is easy to show that $\textrm{Tr}(D_{g}^{\prime}\rho D_{g}^{\prime\dagger})= \frac{1}{N}(g_{00} + (N-1)\lambda^{2})$, so we can write
\begin{eqnarray}
P(\sigma)
\leq
\frac{1}{||D_{g}^{\prime}||^{2}}
\frac{g_{00}^{2} + (N-1)\lambda^{2}}{N}.
    \label{implementationCost8}
\end{eqnarray}
Equation (\ref{implementationCost4}) ensures that the largest possible value $p$ can assume is $p=\frac{1}{||D_{g}^{\prime}||^{2}}$. In addition for the interval $\phi$ can outperform Grover's algorithm we have $||D_{g}^{\prime}||=\lambda$. So we have
\begin{eqnarray}
P(\sigma)
&=&
1 - \frac{1}{N}
\left(
    1 - \frac{g_{00}^{2}}{||D_{g}^{\prime}||^{2}}
\right),
    \label{implemantationCost9}
\end{eqnarray}
or, considering (\ref{metricGrover16}),
\begin{eqnarray}
P(\sigma)
&=&
1 - \frac{1}{N}
\left(
    1 - \tan\theta\cot\phi
\right).
    \label{implementationCost10}
\end{eqnarray}
Since $\frac{g_{00}}{||D_{g}^{\prime}||}< 1$ for our allowed $\phi$ and $P(\sigma)$ represents a probability, then we have that
\begin{eqnarray}
1 -\frac{1}{N}
<
P(\sigma)
<
1.
    \label{implementationCost11}
\end{eqnarray}
So, for sufficient large search space we have high probability that $\mathcal{E}(VU_{f} \rho U_{f}^{\dagger}V^{\dagger})$ implement the non-unitary operation $\mathcal{D}_{g}= \sqrt{p}D_{g}^{\prime}$.

Once $\mathcal{D}_{g}$ is implemented, we will have the state
\begin{eqnarray}
|\varphi\rangle
&=&
\frac{1}{\sqrt{\langle \psi|U_f D_{g} D_{g}^{\dagger} U_{f}|\psi\rangle}
}
D_{g}U_{f}|\psi\rangle
\nonumber
\\
&=&
\frac{1}{\sqrt{
    \textrm{Tr}
        \left(
            D_{g}U_{f}|\psi\rangle\langle \psi| U_{f}^{\dagger} D_{g}^{\dagger}
        \right)
    }
}
D_{g}U_{f}|\psi\rangle,
    \label{implementationCost12}
\end{eqnarray}
where we have used $D_{g}= D_{g}^{\prime}V$. By (\ref{implementationCost8}) and $p= \frac{1}{||D_{g}||^{2}}$ we have
\begin{eqnarray}
|\varphi\rangle
&=&
\frac{\sqrt{p}}{\sqrt{P(\sigma)}}
D_{g}U_{f}|\psi\rangle.
    \label{implementationCost13}
\end{eqnarray}

The conditional probability of measuring a solution $|x\rangle$ once successfully implemented $D_{g}$ is
\begin{eqnarray}
\sum_{x\in S}|\langle x|\varphi\rangle|^{2}
&=&
\frac{p}{P(\sigma)}
\sum_{x\in S}|\langle x| D_{g}U_{f}|\psi\rangle|^{2}.
    \label{implementationCost14}
\end{eqnarray}
Multiplying by $P(\sigma)$ we have the total probability of success $P_{succ}$ of implementing the non-unitary $D_{g}^{\prime}$ and measure a solution $|x\rangle$, 
\begin{eqnarray}
P_{succ}
&=&
p
\sum_{x\in S}
|\langle x| D_{g}U_{f}|\psi\rangle|^{2}.
    \label{implementationCost15}
\end{eqnarray}
As $||D_g^{\prime}||= \sin\phi/\sin\theta$ and $\phi = \pi/2 - \theta/2$, we can use $\sin(\theta)=2\sin(\theta/2)\cos(\theta/2)$ and the definition $\sin(\theta/2) =\sqrt{\frac{M}{N}}$ and write
\begin{eqnarray}
P_{succ}
&=&
4 \frac{M}{N}
\sum_{x\in S}
|\langle x| D_{g}U_{f}|\psi\rangle|^{2}.
    \label{implementationCost16}
\end{eqnarray}
By noting that $D_{g}= g_{00}|0\rangle\langle 0| - \lambda \sum_{y\geq 1| y \not\in S}|y\rangle\langle y| -\lambda\sum_{y\geq 1 | y\in S} |y\rangle\langle y|$ and $U_{f}|\psi\rangle = \frac{1}{\sqrt{N}}\sum_{z\not\in S}|z\rangle - \frac{1}{\sqrt{N}}\sum_{z\in S}|z\rangle$ it is easy to show that
\begin{eqnarray}
\sum_{x\in S}
|\langle x| D_{g}U_{f}|\psi\rangle|^{2}
&=&
\frac{1}{4}
-
\delta_{g(0)0}
\left[
    \frac{N - M(N - 2)}{M(N - 2 M)}
\right],
    \label{implementationCost16}
\end{eqnarray}
where $\delta_{ij}$ is the Kronecker delta, $g(x) = 1 - f(x)$ and we have used $\phi = \pi/2 -\theta/2$. We have that $\delta_{g(0)0} = 0$ if $0\in S$. So, our best probability of success occurs when $0\not\in S$, where it takes the value
\begin{eqnarray}
P_{succ}
&=&
\frac{M}{N}.
    \label{implemantationCost17}
\end{eqnarray}
This is exactly the same probability of sorting a random $x\in \mathbb{Z}_{N}$ such that $x\in S$. In this sense one have to repeat the algorithm $1/P_{succ}$ times to be to guarantee the success of finding a solution to $f(x)=1$. In another words, to obtain success in finding a solution, one has to repeat the algorithm $O(\frac{N}{M})$ times, giving no advantage towards the classical algorithm.

This happens because the process of turning $D_{g}^{\prime}$ a valid Kraus operator transforms it in the normalized operator $\mathcal{D}_{g}$. The quantum operator (\ref{implementationCost5}) ensures the application of $\mathcal{D}_{g}$ with high probability, as we see in (\ref{implementationCost11}). So, the implemented operator it was the normalized $\mathcal{D}_{g}$, not $D_{g}$. This is evident in expression (\ref{implementationCost13}) that we can rewrite as
\begin{eqnarray}
|\varphi\rangle
&=&
\frac{1}{\sqrt{P(\sigma)}}
\mathcal{D}_{g}
|\psi_{f}\rangle,
    \label{implementationCost18}
\end{eqnarray}
where we have used the definition (\ref{implementationCost3}) and $|\psi_{f}\rangle = VU_{f}|\psi\rangle$. In this process, the normalization constant $p$ weights the probability (\ref{implementationCost15}) so that we have to repeat the whole algorithm $O\left(\frac{1}{p}\right)$ times to have a chance of finding some $x\in S$. Using the angle $\phi= \pi/2 -\theta/2$, we have that $\frac{1}{p}= 4\frac{N}{M}$, where the $O\left(\frac{N}{M}\right)$ comes from.

The desired operation we need is
\begin{eqnarray}
D_{g}
&=&
\frac{1}{\sqrt{p}}
\mathcal{D}_{g}.
    \label{implementationCost19}
\end{eqnarray}
This can be archived by implementing a polynomial $P(X)= a X$, where we can identify $X$ with $\mathcal{D}_{g}$ and $a$ with the constant $\frac{1}{\sqrt{p}}$. If we could implement the aforementioned polynomial, then we expected to need less operations to obtain success in measuring a $x\in S$. Nonetheless, Kraus operator approach cannot help us in this task because it is implicit a measurement in an ancilla qubit. That is, for a Kraus operator $K$ it is understood that
\begin{eqnarray}
K
&=&
\mathrm{Tr}_{A}
\left(
    \mathcal{U}\rho\otimes| A\rangle\langle A| \mathcal{U}^{\dagger} 
\right),
    \label{implementationCOst20}
\end{eqnarray}
where $\rho$ is the state of interest, $|A\rangle\langle A|$ is some state in the ancilla register and $\textrm{Tr}_{A}$ is the partial trace with respect with the ancilla system. The operator $\mathcal{U}$ is some unitary that evolves $\rho\otimes |A\rangle\langle A|$ at once. Note that in this approach $\mathcal{U}$ it is not necessary known and what is important is the partial trace $K$ of the evolved entangled system. In the next section, we will embed $\mathcal{D}_{g}$ in a larger unitary operator and make it possible to recover the normalization loss.

\subsection{Block encoding approach}

In this section we will try to recover the normalization loss obtained when we tried to transform $D_{g}$ in the normalized operator $\mathcal{D}_{g}^{\prime}$. Here we will propose a unitary $U$ given by
\begin{eqnarray}
U
&=
\left(
\begin{matrix}
 \mathcal{D}_{g} & * \\
       *         & *
\end{matrix}
\right),
    \label{implementationCost21}
\end{eqnarray}
that must be applied on the state $|\psi_{f}\rangle = VU_{f}|\psi\rangle\otimes |A\rangle$, where $|A\rangle$ is a $1$-qubit ancilla. The $*$ means some block matrix which ensures the unitarity of $U$. One can think of  it as a trying to discover the unitary inside the trace in (\ref{implementationCOst20}). The idea is that after a measure on ancilla system we have the action of $\mathcal{D}_{g}$ up to an error $\varepsilon$, that is
\begin{eqnarray}
(\langle 0|_{A}\otimes I)
U
(|0\rangle_{A}\otimes I)
&=&
\mathcal{D}_{g} + \varepsilon
=
\sqrt{p}D_{g}^{\prime} + \varepsilon
    \label{implementationCost22}
\end{eqnarray}
Nonetheless we will need to apply $D_{g}^{\prime}= \frac{1}{\sqrt{p}}\mathcal{D}_{g}$. For archive this we will choose a $d$ degree Chebyshev polynomial $T_{d}(x) = a x$, for some real constant $a$, to approximating the linear operation $\frac{1}{\sqrt{p}}U$ in the interval $[-\sqrt{p},\sqrt{p}]$ \cite{Gilyen2019}. This requires a degree $d= \frac{1}{\sqrt{p}}$ polynomial and a complexity of $O(d\log||\varepsilon||^{-1})$. As, for our case $\frac{1}{\sqrt{p}}=||D_{g}^{\prime}||$ and the chosen angle $\phi = \pi/2 -\theta/2$ gives $||D_{g}^{\prime}|| = \sqrt{N/M}$, this operation will cost us a complexity of $O\left(\sqrt{\frac{M}{N}}\log||\varepsilon||^{-1}\right)$ and we recover the normalization loss we had in Kraus analysis. 

\section{Conclusions}
\label{conclusions}

We have developed a new view of quantum search algorithm. We first understand each iteration of the standard Grover's algorithm as an application of the oracle $U_{f}$ followed by an application of a correlation onto a dual vector space in respect with the pseudo-metric tensor $\Lambda$, given by (\ref{theProblem4}), transformed by Walsh-Hadamard transform $H^{\otimes n}$. Such perspective allowed us to observe the exactly mechanism that make it possible to rotate the initial state by a fixed angle $\theta$ in direction to the solution subspace. Then we have generalized the Grover's algorithm by proposing a substitution of $\Lambda$ by a general indefinite $2$-rank tensor and choose its elements so that a single iteration of proposed algorithm can rotate the initial state to the space of the solution. Geometrically, it means that we have changed the first iteration resulting vector to a new dual space, that induced by the Walsh-Hadamard transformed diagonal $2$-rank tensor $D_{g}$. This $2$-tensor tensor scales the first component by a factor of $\frac{\cos\phi}{\cos\theta}$ and all the others by a factor of $-\frac{\sin\phi}{\sin\theta}$, where $\phi$ is an arbitrary angle and $\theta$ depends on the size $N$ of search space and the number $M$ os solutions. For $\phi = \pi/2 - \theta/2$ the first component scales to zero and the others scale by a factor of $\frac{1}{4}\sqrt{\frac{M}{N}}$.

From the point of view of quantum algorithm, $D_{g}$ cannot be implemented in a quantum computer because it is a non-unitary operator. In this sense, the advantage of our algorithm in relation to Grover's is just mathematically. When one tries to implement $D_{g}$ in quantum registers via some available technique, one pay the price of using unitary in larger space  and by normalization. This aligns with the proofs that state that Grover's algorithm is optimal.

We analyze the implementation of $D_{g}$ in two different frameworks: Kraus operator and block encoding. Kraus operator approach is not per si one algorithm, this is a physical analyzis in the original Hilbert space. We see that any valid must be trace decreasing or trace preserving, but never trace increasing. Our operation lies in the last one class and what is possible to do is working with the normalized operation. The price of normalization is paid in post selection since the total probability of measuring a solution is of the order of $\frac{M}{N}$ even though the probability of implementing $D_{g}^{\prime}/||D_{g}^{\prime}||$ is high. In this context, it is necessary to repeat the process $O\left(\frac{N}{M}\right)$ times to increase the chance of measuring a solution. This is the same complexity of the classical algorithm of search.

The block encoding approach comprises of embed the normalized operator into left top of a unitary acting on a larger Hilbert space and the probability of success of implementing the desired operation depends on the probability of measuring the correct ancilla. As well as in the Kraus analysis, this probability is dominated by normalization factor. However, as this operation is unitary, than it is possible to implement a $d$ degree Chebyshev polynomial of this unitary that approximate, in the interval $[-||D_{g}^{\prime}||, ||D_{g}^{\prime}||]$, the action of the non-normalized operator $D_{g}^{\prime}$ on the original Hilbert space. This is made by the cost of $O\left(\sqrt{\frac{M}{N}}\log||\varepsilon||^{-1}\right)$. In this sense, block encoding followed by Chebyshev polynomial approximation reaches the Grover's bound for a constant error $\varepsilon$.

It is well known that quantum simulation \cite{MichaelA.Nielsen2010,Zhukov2025,Portugal2018} and quantum walks \cite{Portugal2018,Nahimovs2015} can reproduce the Grover's algorithm results for specific choice of Hamiltonian and coin operator, respectively. In this sense, our algorithm add efforts to bring new perspective of search problem. However, our contribution do not put quantum search into a specific application of a broader theory, it instead generalize Grover's algorithm into a non-unitary extension in which recovers to the original algorithm with a the choice $\phi = \theta$. Up to multiple of $2\pi$ phases, this is the unique choice of $\phi$ that makes $D_{g}$ unitary. We note that we need to know the number of solutions of the search problem in order to build the $D_{g}$, since $\theta = 2\arcsin\sqrt{\frac{M}{N}}$. But we do not consider this a disadvantage compared to Grover's one, because there we need to estimate how many iteration we need to perform. By using Grover's operation and phase estimative, it is possible to estimate the number of solution \cite{MichaelA.Nielsen2010}. But, as we discussed, Grover's operation coincides with our operation for a specific rotation angle. Even though we do not know the number of solutions, for this case, we know that $g = \mathrm{diag}(1,-1,-1,...,-1)$ coincides with $\Lambda$ of the usual Grover's algorithm and can be used in quantum counting, since for this choice of $g$ we have $\mathcal{D}_{g}=\mathcal{D}$ (see Eq. (\ref{theProblem3})). As the complexity of our algorithm using block encoding with a constant error is the same of Grover's, than we can conclude that our only disadvantage is the use of a extra qubit to perform the block encoding.

\end{document}